\documentclass[]{aastex631}
\usepackage{graphicx}
\usepackage{xcolor}

\newcommand{\hi}{\ion{H}{1}}
\newcommand{\nai}{\ion{Na}{1}}

\newcommand{\hei}{\ion{He}{1}}
\newcommand{\caii}{\ion{Ca}{2}}
\newcommand{\oi}{\ion{O}{1}}

\newcommand{\n}{[\ion{N}{2}]}

\newcommand{\av}{A$_V$}

\newcommand{\lsun}{L$_{\odot}$}
\newcommand{\msun}{M$_{\odot}$}
\newcommand{\rsun}{R$_{\odot}$}
\newcommand{\msunyr}{M$_{\odot}$\,yr$^{-1}$}
\newcommand{\macc}{$\dot{M}_{\mathrm{acc}}$}

\newcommand{\lacc}{$L_{\mathrm{acc}}$}
\newcommand{\lacci}{$L_{\mathrm{acc(i)}}$}
\newcommand{\lumi}{$L_{\mathrm{i}}$}
\newcommand{\rstar}{$R_{\mathrm{*}}$}

\newcommand{\lstar}{$L_{\mathrm{*}}$}
\newcommand{\mstar}{$M_{\mathrm{*}}$}
\newcommand{\Mbol}{$M_{\mathrm{bol}}$}
\newcommand{\teff}{$T_\mathrm{eff}$}

\newcommand{\lbol}{$L_{\mathrm{bol}}$}

\newcommand{\f}{\scriptsize}

\shorttitle{Gaia23bab: a new EXor}
\shortauthors{Giannini et al.}

\graphicspath{{./}{figures/}}

\begin{document}

\title{Gaia23bab : a new EXor}

%\correspondingauthor{Teresa Giannini}
%\email{teresa.giannini@inaf.it}

\author[0000-0002-7035-8513]{T. Giannini}
\affiliation{INAF - Osservatorio Astronomico di Roma, Via di Frascati, 33, 00078, Monte Porzio Catone, Italy}
\author[0000-0003-1560-3958]{E. Schisano}
\affiliation{INAF - Istituto di Astrofisica e Planetologia Spaziali,  Via Fosso del Cavaliere 100, I-00133 Roma, Italy}
\author[0000-0002-9190-0113]{B. Nisini}
\affiliation{INAF - Osservatorio Astronomico di Roma, Via di Frascati, 33, 00078, Monte Porzio Catone, Italy}
\author[0000-0001-6015-646X]{P. {\'A}brah{\'a}m}
\affiliation{HUN-REN Research Centre for Astronomy and Earth Sciences, Konkoly Observatory, Konkoly Thege Mikl\'os \'ut 15-17., H-1121, Hungary}
\affiliation{CSFK, MTA Centre of Excellence, Budapest, Konkoly Thege Mikl\'os \'ut 15-17., H-1121, Hungary}
\affiliation{Department of Astrophysics, University of Vienna, Türkenschanzstrasse 17, A-1180 Vienna, Austria}
\affiliation{Institute of Physics, ELTE Eötvös Loránd University, Pázmány Péter sétány 1/A, 1117 Budapest, Hungary}
\author[0000-0002-0666-3847]{S. Antoniucci}
\affiliation{INAF - Osservatorio Astronomico di Roma, Via di Frascati, 33, 00078, Monte Porzio Catone, Italy}
\author[0000-0002-1892-2180]{K. Biazzo}
\affiliation{INAF - Osservatorio Astronomico di Roma, Via di Frascati, 33, 00078, Monte Porzio Catone, Italy}
\author[0000-0002-4283-2185]{F. Cruz-S{\'a}enz de Miera}
\affiliation{HUN-REN Research Centre for Astronomy and Earth Sciences, Konkoly Observatory, Konkoly Thege Mikl\'os \'ut 15-17., H-1121,
Hungary}
\affiliation{CSFK, MTA Centre of Excellence, Budapest, Konkoly Thege Mikl\'os \'ut 15-17., H-1121, Hungary}
\affiliation{Institut de Recherche en Astrophysique et Plan\'etologie, Universit\'e de Toulouse, UT3-PS, CNRS, CNES, 9 av. du Colonel Roche, 31028 Toulouse Cedex 4, France}
\author[0000-0002-5261-6216]{E. Fiorellino}
\affiliation{INAF-Osservatorio Astronomico di Capodimonte, via Moiariello 16, I-80131 Napoli, Italy}
\author[0000-0002-8364-7795]{M. Gangi}
\affiliation{INAF - Osservatorio Astronomico di Roma, Via di Frascati, 33, 00078, Monte Porzio Catone, Italy}
\affiliation{ASI, Italian Space Agency, Via del Politecnico snc, 00133, Rome, Italy}
\author[0000-0001-7157-6275]{A. K{\'o}sp{\'a}l}
\affiliation{HUN-REN Research Centre for Astronomy and Earth Sciences, Konkoly Observatory, Konkoly Thege Mikl\'os \'ut 15-17., H-1121, Hungary}
\affiliation{CSFK, MTA Centre of Excellence, Budapest, Konkoly Thege Mikl\'os \'ut 15-17., H-1121, Hungary}
\affiliation{Institute of Physics, ELTE Eötvös Loránd University, Pázmány Péter sétány 1/A, 1117 Budapest, Hungary}
\affiliation{Max Planck Institute for Astronomy, K\"onigstuhl 17, 69117 Heidelberg, Germany}
\author[0000-0002-0631-7514]{M. Kuhn}
\affiliation{Centre for Astrophysics Research, University of Hertfordshire, College Lane, Hatfield, AL10 9AB, UK}
\author[0000-0002-6894-1267]{E. Marini}
\affiliation{INAF - Osservatorio Astronomico di Roma, Via di Frascati, 33, 00078, Monte Porzio Catone, Italy}
\author[0000-0002-3632-1194]{Z. Nagy}
\affiliation{CSFK, MTA Centre of Excellence, Budapest, Konkoly Thege Mikl\'os \'ut 15-17., H-1121, Hungary}
\affiliation{HUN-REN Research Centre for Astronomy and Earth Sciences, Konkoly Observatory, Konkoly Thege Mikl\'os \'ut 15-17., H-1121, Hungary}
\author[0000-0002-7409-8114]{D. Paris}
\affiliation{INAF - Osservatorio Astronomico di Roma, Via di Frascati, 33, 00078, Monte Porzio Catone, Italy}

\begin{abstract}
  On March 6 2023, the Gaia telescope has alerted a 2-magnitude burst from 
  Gaia23bab, a Young Stellar Object in the Galactic plane. 
We observed Gaia23bab with the Large Binocular Telescope obtaining optical and near-infrared spectra close in time to the peak of the burst, and collected all public multi-band photometry to reconstruct the historical light curve.
This latter shows three bursts in ten years (2013, 2017 and 2023), whose duration and amplitude are typical of EXor variables.
 We estimate that, due to the bursts, the mass accumulated on the star is about twice greater than if the source had remained quiescent for the same period of time.
Photometric analysis indicates that Gaia23bab is a Class\,II source with age $\la$ 1 Myr, spectral type G3$-$K0,
stellar luminosity $\sim$ 4.0 \lsun\,, and mass $\sim$ 1.6 \msun\,.
The optical/near infrared spectrum is rich in emission lines.
From the analysis of these lines we measured the accretion luminosity and the mass accretion rate (\lacc$^{burst}$\, $\sim$ 3.7 \lsun\,, \macc$^{burst}$\, $\sim$ 2.0\,$\times$\,10 $^{ -7}$ \msunyr)\, consistent with those of EXors.
More generally, we derive the relationships between accretion and stellar parameters in a sample of EXors. We find that, when in burst, the accretion parameters become almost independent of the stellar parameters and that EXors, even in quiescence, are more efficient  than classical T Tauri stars in assembling mass.
\end{abstract}


\section{Introduction}\label{sec:sec1}
Photometric variability is a common feature of low mass ($<$ 2\,\msun), young stellar objects (YSOs, e.g. Megeath et al. 2012).
Very different timescales are involved in the observed 
variability: from short term events (from minutes to days) due
to magnetic 
activity, like surface spots, stellar flares and coronal mass ejections, up 
to long term events (from months to years and centuries) 
induced by extinction changes due to inner-disk warps, or 
abrupt variations in the accretion rate. To this latter class 
of variability events belong the so-called eruptive young 
stars (EYSs), historically categorized as FU Orionis–type 
objects or FUors (Herbig 1977), and EX 
Lupi–type objects, or EXors (Herbig 1989). FUors present 
powerful outbursts of 3\,$-$\,6 mag in the visual band that 
last from several years to centuries, and take from months to
years to reach the peak. Their spectral type depends on the 
observed wavelength, being of F$-$G type in the optical to K$-$M 
type in the near-infrared (NIR; Hartmann \& Kenyon 1996; Audard et al. 2014; Connelley \& Reipurth 2018; Fischer et al. 
2023). During the outburst, the accretion rate is of the order of
10$^{-4}$$-$10$^{-5}$
\msunyr\, and the spectra are dominated by absorption lines. EXor 
bursts have amplitudes of 1$-$3 magnitudes in the optical, last for a few
months or a year, and are recurring (Fischer et al. 
2023).  Their spectra 
resemble those of K$-$ or M$-$type dwarfs rich of emission lines,
showing accretion rates in
burst of the order of 10$^{-6}$$-$10$^{-7}$
\msunyr.

A dozen of EXors were classified by Herbig (1989), due to their resemblance with the prototype of the class EX Lupi. Since then and until the early 2000s, the number of EYSs candidates has just slightly increased to some tens (Audard et al. 2014) and therefore, eruptive accretion episodes have been considered as peculiar and rare events. This view, however, is now rapidly changing thanks to the discoveries of the Gaia telescope, which in eight years of operations has issued alerts\footnote{http://gsaweb.ast.cam.ac.uk/alerts/alertsindex} (Hodgkin et al. 2021) of significant photometric changes in the light curve of about 700 known or candidate YSOs. It is interesting to note that the vast majority of the alerted sources undergo photometric variations that do not fit into the two classical EYS categories, while only a few present light curves resembling those of confirmed EXor or FUor variables. Indeed, just three sources have been claimed as {\it bona-fide} FUors\,: Gaia17bpi (Hillenbrand
et al. 2018), Gaia18dvy (Szegedi-Elek et al. 2020), and Gaia21elv  (Nagy et al. 2023). To these add four further sources that share properties of FUors and EXors\,: Gaia19ajj (Hillenbrand et al. 2019), Gaia19bey (Hodapp
et al. 2020), Gaia21bty (Siwak et al. 2023), and Gaia18cjb  (Fiorellino et al. 2024, submitted),  and four
EXors\,: Gaia18dvz (ESO-H$\alpha$99, Hodapp et al. 2019); Gaia20eae
(Hankins et al. 2020, Ghosh et al. 2022, Cruz-S{\'a}enz de Miera et al. 2022); Gaia19fct (Miller et al. 2015; Park et al.
2022); and Gaia22dtk (Kuhn et al. 2022). 

\begin{figure}[ht!]
\plotone{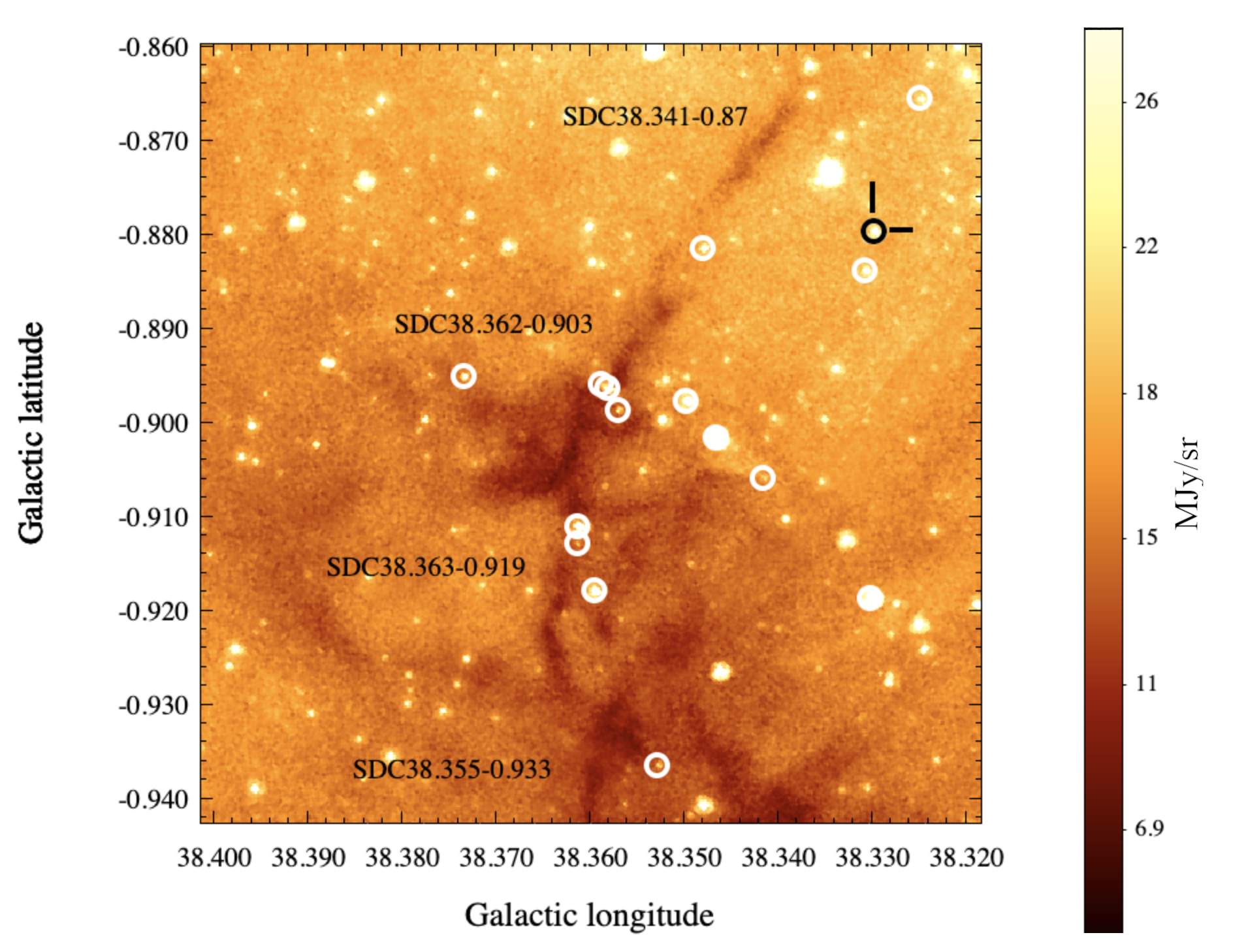}
\caption{IRAC band 4 
(8.0\,$\mu$m) image of a 3$\times$3 arcmin$^2$ 
sky area containing Gaia23bab (marked in black). White circles are the YSOs of the cluster G38.3-0.9, and the labels indicate the location of the Spitzer Dark Clouds (SDC) from Peretto \& Fuller (2009) catalog. The surface brightness scale is given in MJy/sr in the right bar. 
\label{fig:fig1}}
\end{figure}
On 2023 March 6, Gaia alerted a 2-mag burst of
Gaia23bab ($\alpha_{J2000.0}$=19$^h$ 04$^m$ 
26$^s$.68, $\delta_{J2000.0}$=$+$04$^{\circ}$ 
23$^{\prime}$ 57$\farcs$37). Also known as SPICY
97589 (Kuhn et al. 2021), Gaia23bab is a YSO belonging to the small cluster of 
$\sim$ 30 members named G38.3-0.9.
In Figure\,\ref{fig:fig1} we show the 
3\,$\times$\,3 arcmin$^2$ IRAC band 4 image of the 
cluster, where are also four Spitzer 
Dark Clouds (Peretto \& Fuller 2009), within which 
the cluster is partially embedded. 
The membership of Gaia23bab to G38.3-0.9 has been 
demonstrated by Kuhn et al. (2023) on the base of
the parallaxes and proper motions of six members 
of the cluster reported in the Gaia DR3 archive 
(Gaia Collaboration et al. 2016, 2021). The 
cluster parallax is 1.114 $\pm$ 0.056 mas, 
corresponding to a distance  $d$\,=\,900 $\pm$ 45 pc.

The photometric and spectroscopic properties of Gaia23bab are the subject of the present paper. 
We first describe the spectroscopic observations (Sect.\,\ref{sec:sec2}).  Then, the photometric and spectroscopic analysis is presented in Sect.\,\ref{sec:sec3}-\ref{sec:sec4}. 
A short discussion in given in Sect\,\ref{sec:sec5}, while the summary is presented in Sect.\,\ref{sec:sec6}. 

\section{Observations and data reduction}\label{sec:sec2}

\begin{figure}[ht!]
\plotone{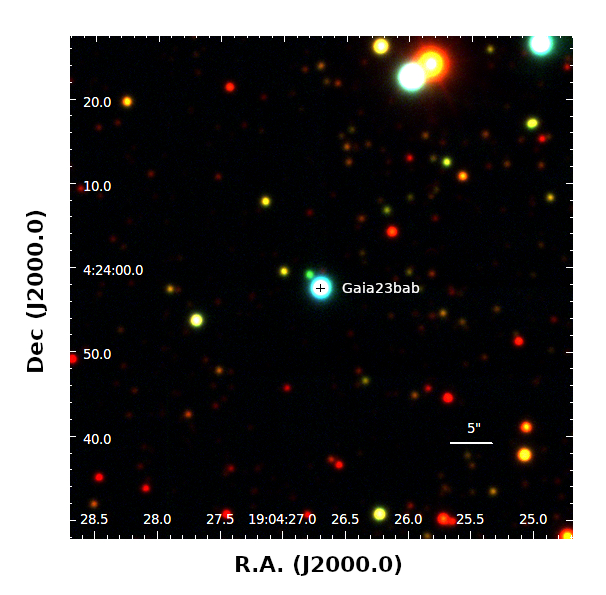}
\caption{{LBT 3-color image of 1$\times$1 arcmin$^2$ sky area centered on Gaia23bab. The color code is the following. Red: $J$+$H$+$K_s$ (LUCI); green: $i$+$z$ (MODS); blue: $g$+$r$ (MODS). Note that Gaia23bab is among the bluest objects in the field.}
\label{fig:fig2}} 
\end{figure}
We observed Gaia23bab  with the 8.4m Large Binocular Telescope (LBT) located at Mount Graham, (Arizona, USA). The long-slit optical spectrum was obtained combining the observations collected on 2023 May 28 and 29, with the Multi-Object Double Spectrograph (MODS, Pogge et al. 2010). 
MODS observations were done with the dual grating mode (Blue + Red channels, spectral range 350$-$950 nm)  by using a 0$\farcs$80 slit ($\Re 
\sim$ 1500 and 1800 in the Blue and Red channels, respectively). The total integration time was 1800 s. The slit 
angle matched the parallactic angle to minimize the wavelength dependence of
the slit transmission.
During the two following nights (2023 May 30 and June 1) the 
LBT Utility Camera in the Infrared (LUCI, Seifert et al. 2003)  was used with the $zJ$ and $HK$ grisms to obtain the 1.0$-$2.4 $\mu$m spectrum. We used 
the G200 low-resolution grating coupled 
with the 0$\farcs$75 slit, corresponding to $\Re \sim$1500.
The observations were performed adopting the standard ABBA 
technique with a total integration time of 1350 s.

Data reduction was done using the Spectroscopic Interactive Pipeline and 
Graphical Interface  (SIPGI, Gargiulo et al. 2022), specifically developed to 
reduce the LBT long-slit spectra.
Data reduction steps of each  MODS spectral-image are: correction for dark and bias, bad-pixel mapping, flat-fielding, correction for optical distortions in the spatial and spectral directions, and extraction of the one-dimensional 
spectrum by integrating the stellar trace along the spatial direction. 
Spectra of arc lamps have been used for wavelength calibration. 
Images in the $griz$ bands  (spatial scale 0\farcs12/px) were obtained to derive the optical photometry of Gaia23bab and calibrate the MODS spectrum, by taking as references all the stars in the field present in the Pan-STARRS  (Panoramic 
Survey Telescope \& Rapid Response System)\footnote{https://outerspace.stsci.edu/display/PANSTARRS/} catalog (Chambers, et al.\ 2016). We estimate $g$=18.97$\pm$0.09 mag, $r$=17.11$\pm$0.09 mag, $i$=15.51$\pm$0.07 mag, $z$=14.74$\pm$0.07 mag.
The inter-calibration between Blue and Red spectral segments was verified by matching the spectral range between 5300 \AA\, and 5900 \AA\, in common between the two channels. 

The raw LUCI spectral images were corrected for bad pixels, flat-fielded, sky-subtracted, and corrected for optical distortions in both the spatial and spectral directions. The telluric lines present in the final spectrum were removed by dividing the target spectrum by that of a spectro-photometric standard star observed immediately after the target and corrected for its intrinsic \hi\, recombination lines in absorption. Wavelength calibration was obtained from arc lamps spectra. $J$, $H$ and $K_s$ images of Gaia23bab  (spatial scale 0\farcs12/px) were acquired to flux calibrate the $zJ$ and $HK$ segments. The photometry was computed based on the 2MASS\footnote{https://irsa.ipac.caltech.edu/Missions/2mass.html}  magnitudes of the sources present in the image field. We obtain $J$=13.2$\pm$0.1 mag, $H$=12.24$\pm$0.07 mag, $K_s$=11.47$\pm$0.07 mag. In Figure\,\ref{fig:fig2} we show the composite image of a 1$\times$1 arcmin$^2$ sky area around Gaia23bab obtained from all the LBT images, where MODS $g$ and $r$ images are in blue, MODS $i$ and $z$ images are in green, and LUCI $J$, $H$ and $K_s$ images are in red. Noticeably, Gaia23bab is among the bluest objects in the field.

\section{Photometric analysis}\label{sec:sec3}
\subsection{Light curve}\label{sec:sec3.1}

\begin{figure}[ht!]
\includegraphics[width=1.\columnwidth]{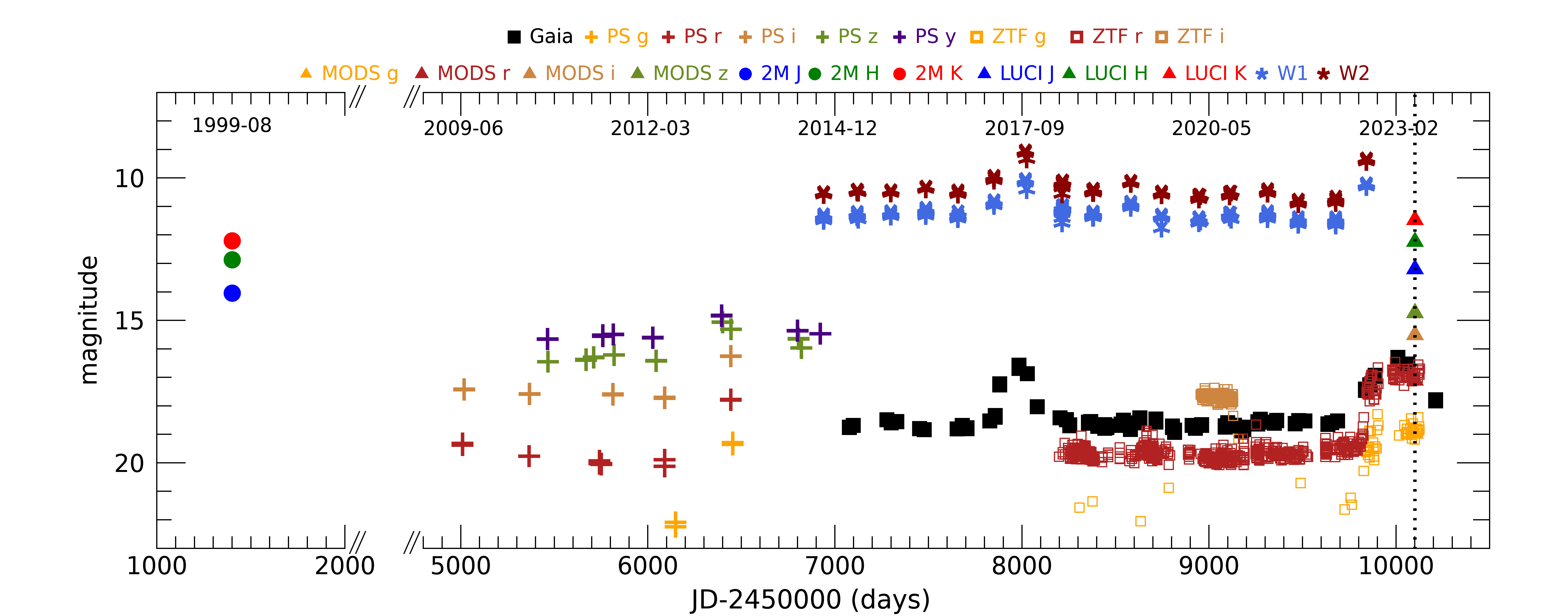}
\caption{Light curve of Gaia23bab. Different symbols are the photometric points obtained within different surveys (filled squares: Gaia ($G$), crosses: Pan-STARRS; open squares: ZTF, filled circles: 2MASS, asterisks: NEOWISE, filled triangles: LBT). Different colors indicate different filters, as indicated on top. The calendar date is indicated as well. The dotted black line marks the date of the LBT spectroscopy.   
\label{fig:fig3}}
\end{figure}

Figure\,\ref{fig:fig3} shows the light curve of 
Gaia23bab during the last fourteen years. Between 2009 and 2014 the optical photometry in the $grizy$ bands was obtained within the Pan-STARRS survey, with a typical sampling of one or two points per year. The Pan-STARRS data, and in particular those in $z$ and $y$ bands, reveal that a burst has 
occurred before the advent of Gaia, roughly between 2012 April and 2014 June.
A more continuous monitoring has been performed
since 2014 by Gaia and, starting three years later, by the survey 
ZTF\footnote{https://www.ztf.caltech.edu} (Zwicky Transient 
Facility), which covers the $gri$ bands. As already noted by Kuhn et al. (2023) the Gaia light curve shows a $\sim$ 
2 mag burst 
between April and November 2017 that did not triggered a Gaia alert. Then, Gaia23bab remained quiescent for about six years.
During this period the photometric points in the $G$ and $r$-ZTF bands differ by approximately 1 mag, with a difference from what is observed in other sources, where they roughly coincide (e.g. Cruz-S{\'a}enz de Miera et al. 2022, Nagy et al. 2023). The offset reduces and finally disappears when Gaia23bab starts to brighten.  The two filters have similar $\lambda_{ref}$ (6251.5 \AA\, and 6201.2 \AA\,, respectively), but the $G$ filter is broader ($G$-FWHM\,=\,4396.69 \AA\,, $r$-FWHM\,=\,1397.73 \AA\,)\footnote{https://svo.cab.inta-csic.es/main/index.php} and it extends much farther toward longer wavelengths. 
This suggests that if Gaia23bab  is redder when fainter, it may not decrease as much in $G$ as in $r$-ZTF. Alternatively, the same behaviour in the light curve is expected if the $G$ flux of Gaia23bab is contaminated by that of a nearby source of similar magnitude, whose contribution becomes increasingly negligible during the burst. The ability of Gaia to resolve a nearby double star of equal luminosity is 0\farcs{23} in the along-scan and 0\farcs{70} in the across-scan direction, independent of the brightness of the primary (de Bruijne et al. 2015). This is higher than  the ZTF's spatial sampling\footnote{https://www.ztf.caltech.edu/ztf-camera.html} of 1\farcs{0}/px, and therefore it is unlikely that the $G$ flux is contaminated while the $r$-ZTF is not. Furthermore, in both MODS and LUCI images the star closest to Gaia23bab is separated by about 2\farcs0 (see Figure\,\ref{fig:fig2}).

The last optical points in the light curve are the MODS $grzi$ magnitudes. Our observations have been obtained at the beginning of the declining phase, when the $r$-ZTF magnitude was about 0.5 mag fainter than that at the burst peak.

In the infrared, Gaia23bab has been monitored since 2014 within the NEOWISE\footnote{https://www.nasa.gov/mission$\_$pages/WISE/main/index.html} survey
in the W1 and W2 bands at 3.4 $\mu$m and 4.6 $\mu$m. The last two bursts have been registered as about 1.8 
mag brightening in both bands.  In addition, in Figure\,\ref{fig:fig3} we plot
the $JHK_s$ 
2MASS
and LUCI photometry, to compare the magnitudes in 
quiescence and burst. Their difference is $\Delta J$\,=0.85 mag, $\Delta H$\,=0.64 mag, $\Delta K$\,=0.74 mag.

In Table\,\ref{tab:tab1} we summarize
the light curve properties of the three bursts, which appear remarkably similar. First, the amplitude is always comparable to each burst to other ($\sim$ 2 mag in $r$/$G$ and $\sim$ 3 mag in $g$). Second, the time elapsed between the first and the second burst is approximately 50 months, i.e. roughly that between the second and the third (66 months). This frequency, together with the lack of periodicity, has been already observed in EXors (e.g. V1118Ori, Giannini et al. 2020; ASASSN-13db, Sicilia-Aguilar et al. 2017; Gaia19fct, Park et al. 2022). Finally, we have evaluated the burst rising (declining) speed in each band by fitting with straight lines the rising and declining data in the light curve. In all cases the rising speed is of some thousandths of magnitude per day and slightly faster than the declining speed. These values are similar to those observed in V1118 Ori (Giannini et al. 2020), in Gaia20eae (Cruz-S{\'a}enz de Miera et al.\,2022), and in V2492 Cyg (Hillenbrand et al. 2013, Giannini et al. 2018).

\begin{deluxetable*}{ccccccc}
\tablecaption{Features of the bursts of Gaia23bab. \label{tab:tab1}}
\tablewidth{0pt}
\tablehead{
\colhead{Burst ID} & \colhead{Peak date}  & \colhead{Instr-band} & \colhead{$\Delta$mag}& \colhead{Duration}& \colhead{Rising speed}& \colhead{Declining speed} \\
\colhead{}        & \colhead{(calendar date)} &  \colhead{}  & \colhead{(mag)} & \colhead{(months)} & \colhead{(mmag/d)}& \colhead{(mmag/d)}} 
\startdata
1& 2013 Jun & Pan-STARRS-$g$& 2.8            & $>$11        &  9   & -     \\
1& 2013 Jun & Pan-STARRS-$r$& 2.1            & $>$12        &  6   & -     \\
1& 2013 Jun & Pan-STARRS-$i$& 1.5            & $>$12        &  4   & -     \\
1& 2013 Jun & Pan-STARRS-$z$& 1.1            & 25        &  3   & 2     \\
1& 2013 May & Pan-STARRS-$y$& 0.8            & 26        &  3   & 2     \\
2& 2017 Aug & Gaia-$G$& 1.8                  & 8         &  14  & $>$9   \\
3& 2023 Mar & Gaia-$G$& 2.0                  & $\sim$12.5&  7   & $<$ 7  \\
3& -        & ZTF-$g$& 2.8                   & $>$10     &  5   & -     \\
3& 2023 Feb & ZTF-$r$& 2.0                   & $>$10      & 6   & $<$ 4  \\
\enddata
\end{deluxetable*}

\subsection{Color-color diagrams}\label{sec:sec3.2}

\begin{figure}
\begin{center}
\includegraphics[width=1.\columnwidth]{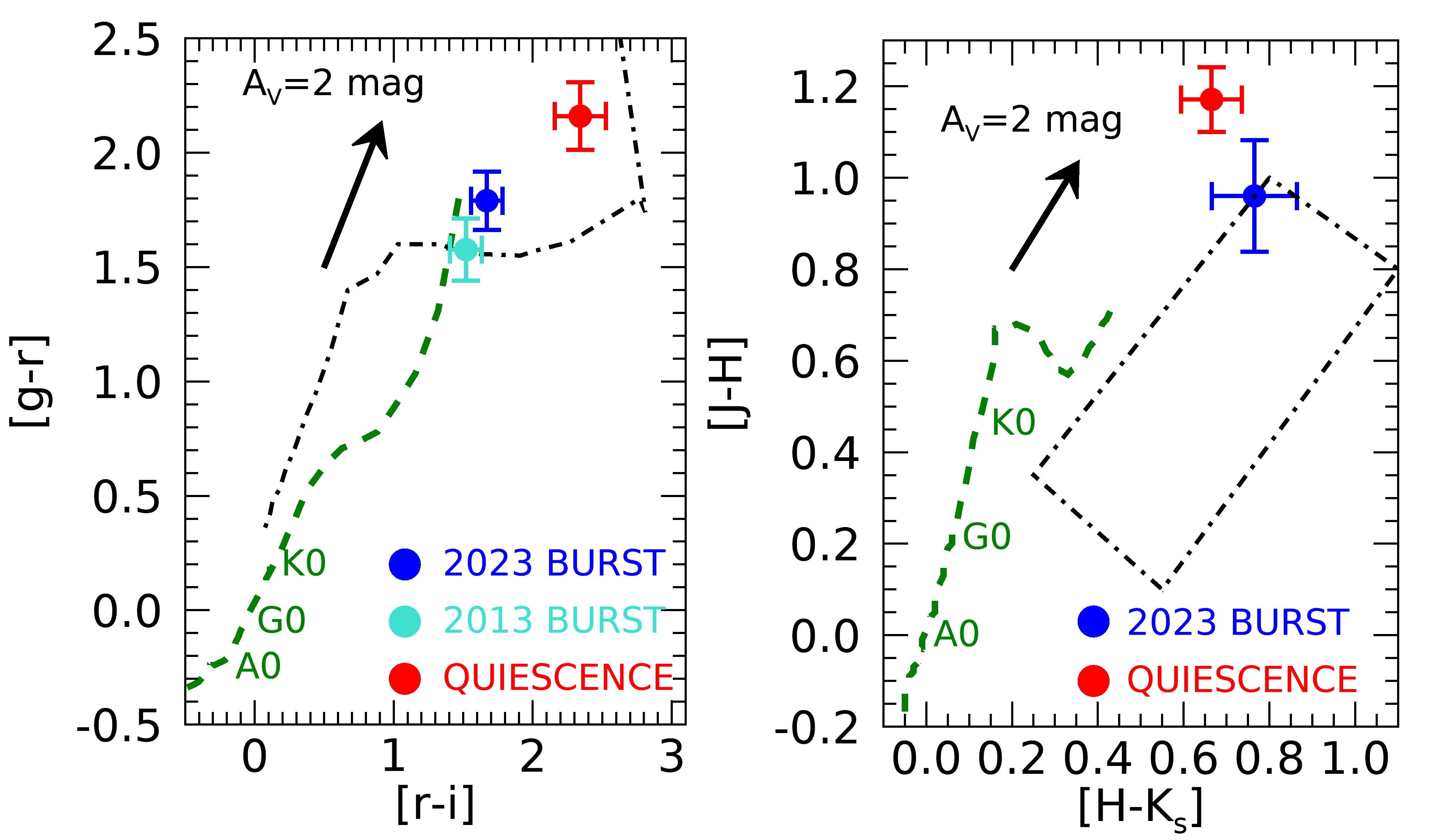}
\end{center}
\caption{\label{fig:fig4} Left panel: two-color optical plot [$g-r$] versus [$r-i$]. Red and blue/cyan dots indicate quiescent and burst points (LBT/Pan-STARRS data). The black dashed-dotted line is the {\it locus} of young stars of 400$-$600 Myr (Kraus \& Hillenbrand 2007), while the green dashed line represents the main-sequence stars with the positions of the A0, G0 and K0 spectral type stars indicated. The arrow represents the direction of the extinction vector corresponding to A$_V$ = 2 mag (reddening law of Cardelli et al. 1989). Right
panel: two-color near-infrared plot [$J-H$]  vs. [$H-K$]. Quiescent (2MASS) and burst (LUCI) data are colored in red and blue, respectively. The dashed-dotted black rectangle is the {\it locus} of the HAeBe stars (Hern{\'a}ndez et al. 2005), while the green dashed line represents the main-sequence stars with the positions of the A0, G0 and K0 spectral type stars indicated. The arrow represents the direction of the extinction vector corresponding to A$_V$ = 2 mag (reddening law of Cardelli et al. 1989).}
\end{figure}

To investigate the nature of the photometric variability of 
Gaia23bab, we plot in Figure\,\ref{fig:fig4} the optical [$g-r$] vs. [$r-i$] and near-infrared [$J-H$] vs. [$H-K_s$] color-color diagrams for
burst and quiescence phases. In the left panel, we show the optical colors during the 2013 (Pan-STARRS, cyan dot) and 2023 (LBT, blue dot) bursts, while the quiescence colors (red dot) have been derived 
by averaging the optical photometries between 2009 and
2013.  We note that the two bursts have similar colors, being both bluer with respect to quiescence of about 0.4/0.6 mag in [$g-r$] and 0.7/0.8 mag in  [$r-i$]. This effect is often seen as a consequence of the dust clearing during the burst (e.g. Hillenbrand et al. 2019) and the increasing contribution to the accretion luminosity at UV wavelengths (e.g. Venuti et al. 2014). 
The colors of 400$-$600 Myr young stars (Kraus \& Hillenbrand 2007) and those of main-sequence stars are also plotted. 
Considering that the stellar spectral type is  G$-$K (Sect.\,\ref{sec:sec3.4}) we estimate an extinction of 5$-$6 magnitudes during bursts and $\sim$ 8 mag in quiescence.
In the right panel, the blue/red dots are the LUCI and 2MASS photometries, obtained more than 20 years apart. The rectangle is the {\it locus} of un-reddened Herbig AeBe (HAeBe) stars of spectral type B$-$F (Hern{\'a}ndez et al. 2005). Considering the error bars, the NIR colors of Gaia23bab marginally fall in this {\it locus}\, if de-reddened by A$_V$\,$\sim$\, 5$-$6 mag in burst. Compared to quiescence, the colors of the burst are $\sim$ 0.2 mag bluer in [$J-H$] and equal within the errors in [$H-K_s$], probably because only the internal regions of the disk are significantly heated by the burst event. This view is also supported by the approximate equality of NEOWISE colors between quiescence and burst, being [W1-W2]$^{quiesc}$ = 0.27$\pm$0.03 mag and [W1-W2]$^{burst}$ = 0.32$\pm$0.03 mag.

\subsection{Spectral Energy Distribution}\label{sec:sec3.3}

\begin{figure}[ht!]
\plotone{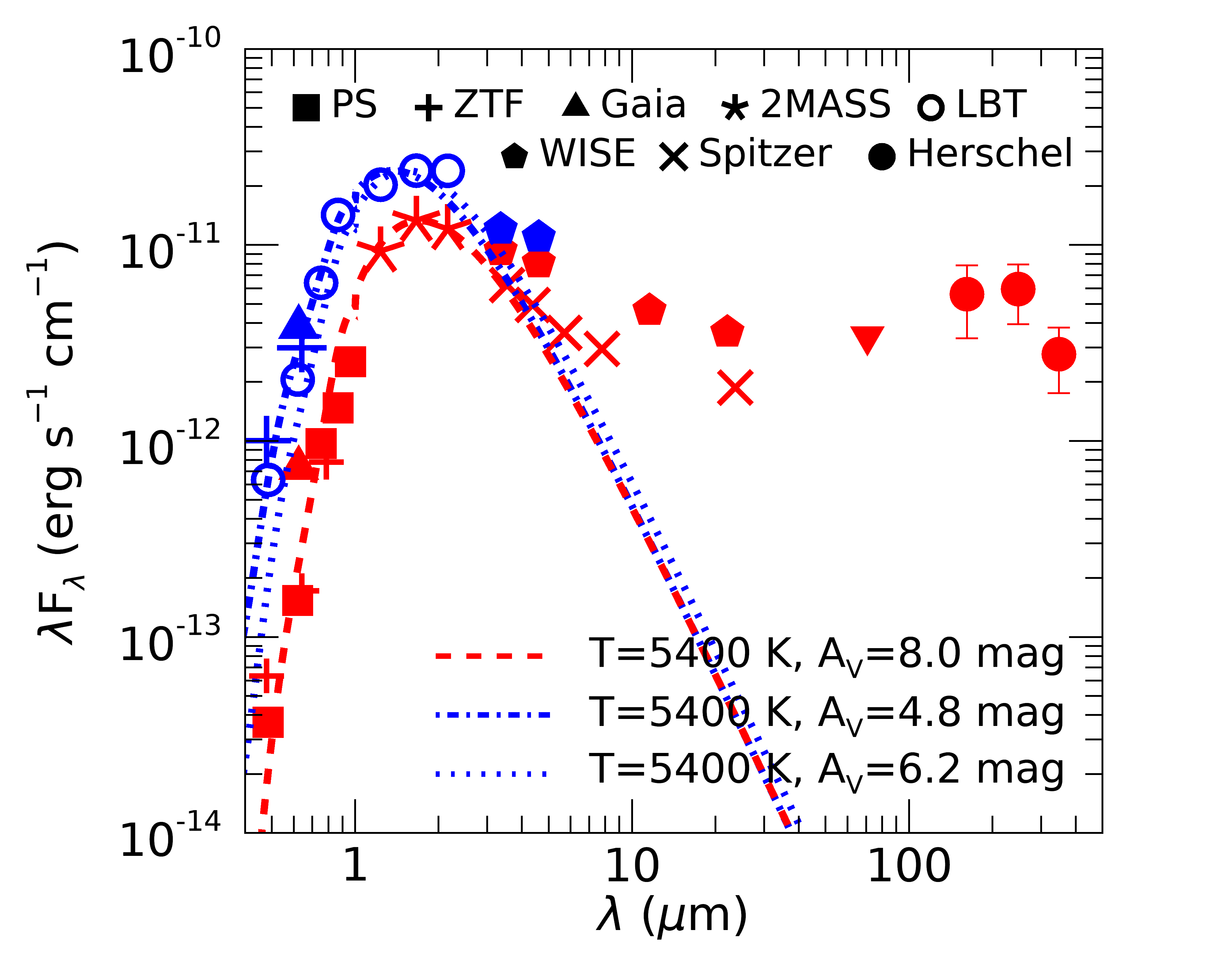}
\caption{Spectral Energy Distribution (SED) of Gaia23bab using data at different epochs. In red and blue are shown the data (not corrected for extinction) in quiescence and in burst, respectively. The down triangle is the 2$\sigma$ upper limit at 70\,$\mu$m. Photometric points from different catalogs are plotted with different symbols. The red and blue dashed lines are the black-body function at T\,=\,5400\,K, 
extincted for A$_V$\,=\,8.0 mag and 6.1 mag, respectively (see Sect.\ref{sec:sec3.2}, Sect.\ref{sec:sec3.4}, and Sect.\ref{sec:sec4.1}).
\label{fig:fig5}}
\end{figure}

In addition to the data shown in the light curve, mid-infrared photometry of Gaia23bab is present in the WISE survey catalogs (AllWISE at 3.4, 4.6, 12.0, and 22.0 $\mu$m, and WISE Post-Cryo Database at 3.4 and 4.6 $\mu$m), and in the Galactic Legacy Infrared Midplane Survey Extraordinaire (GLIMPSE\footnote{http://irsa.ipac.caltech.edu/data/SPITZER/GLIMPSE/}), which contains Spitzer IRAC photometry at 3.6, 4.5, 5.8, and 8.0 $\mu$m. The Spitzer MIPS flux at 24 $\mu$m was taken during the MIPSGAL\footnote{http://mipsgal.ipac.caltech.edu/} survey and retrieved from the Gutermuth \& Heyer (2014) catalog.  
We note that the AllWISE fluxes in W1, W2, and W4 bands are between 0.3 and 0.6  magnitudes brighter than the Spitzer magnitudes in the similar  IRAC1, IRAC2, and MIPS1 bands. In principle, this could be explained by the presence of sources near Gaia23bab falling in the large WISE beam. In the WISE co-added images, however, the closest star is $\sim$ 15$^{\prime\prime}$ apart in the SE direction, and thus it might slightly contaminate the Gaia23bab flux only in W4 band, where the FHWM of the WISE beam is 12$^{\prime\prime}$. A more likely explanation is that at the time of the WISE cryogenic survey (December 2009$-$August 2010), Gaia23bab was at a higher brightness level than in quiescence. This is indeed confirmed by the magnitudes in the W1 and W2 bands obtained in October 2010 during the post-cryo survey, which are all about 0.2$-$0.3 magnitudes fainter than the AllWISE data. While this result is not enough to conclude that a burst occurred between 2009 and 2010, it is certainly  evidence of Gaia23bab's mid-infrared variability. 

In the far-infrared, Gaia23bab was observed within the Infrared Galactic Plane Survey (Hi-GAL) Herschel key-program (Molinari et al. 2010) that surveyed the Galactic Plane with the photometers PACS\footnote{https://www.cosmos.esa.int/web/herschel/pacs-overview} (Photodetector Array Camera and
Spectrometer) and SPIRE\footnote{https://www.cosmos.esa.int/web/herschel/spire-overview} (Spectral and Photometric Imaging REceiver) at 70, 160, 250, 350, and 500\,$\mu$m. We derived the Herschel magnitudes applying both aperture and gaussian photometry, getting reliable measurements at 160, 250, and 350\,$\mu$m. At 70\,$\mu$m the source is barely visible, so that we were only able to estimate an upper limit to the flux, while at 500\,$\mu$m Gaia23bab is confused within the strong background emission. 
The Spectral Energy Distribution (SED) is presented in Figure\,\ref{fig:fig5}. In red and blue are shown the observed photometric
points of quiescence and burst\footnote{Hereinafter, with 'burst' we will indicate the beginning of the declining phase of the 2023 burst.}, respectively. The Pan-STARRS and 
ZTF points are the average values of the data in the light curve for each 
state. In the mid-infrared, we have taken the Spitzer and WISE photometries as representative of the quiescence and burst phase, respectively. Furthermore, considering 
that the amplitude of the variability decreases with $\lambda$, for $\lambda$\,$>$\,24 $\mu$m we have adopted the same 
photometric points for computing the bolometric luminosity (\lbol) in quiescence and in burst. 

We have first de-reddened the photometries up to 24\,$\mu$m with \av\,$\sim$\,8\,mag in quiescence and \av\, between 4.8\,mag and 6.2\,mag in burst, based on the  estimates of the following spectroscopic analysis (Sect.\ref{sec:sec4.1}).
\lbol\, has then been derived integrating the area in the plane F$_\lambda$\,vs.\,$\lambda$ by interpolating with 
straight lines the SED data. A bolometric correction was applied to take
into account the contribution at $\lambda$ $>$ 350\,$\mu$m, considering that 
the emission decreases as 1/$\lambda^2$.
Assuming \lbol\,=\,\lstar\,+\lacc\, (\lstar\, and \lacc\, being the stellar and the accretion luminosity, respectively) and $d$\,=\,900\,pc, we obtain
\lbol$^{quiesc}$\,=\,\lstar\,+\lacc$^{quiesc}$\,$\sim$\,5\,\lsun\, and \lbol$^{burst}$\,=\,\lstar\,+\lacc$^{burst}$\,=\,10.3$\pm$3.9\,\lsun\,.

\subsection{Stellar properties}\label{sec:sec3.4}

\begin{figure}
\begin{center}
\includegraphics[width=1\columnwidth]{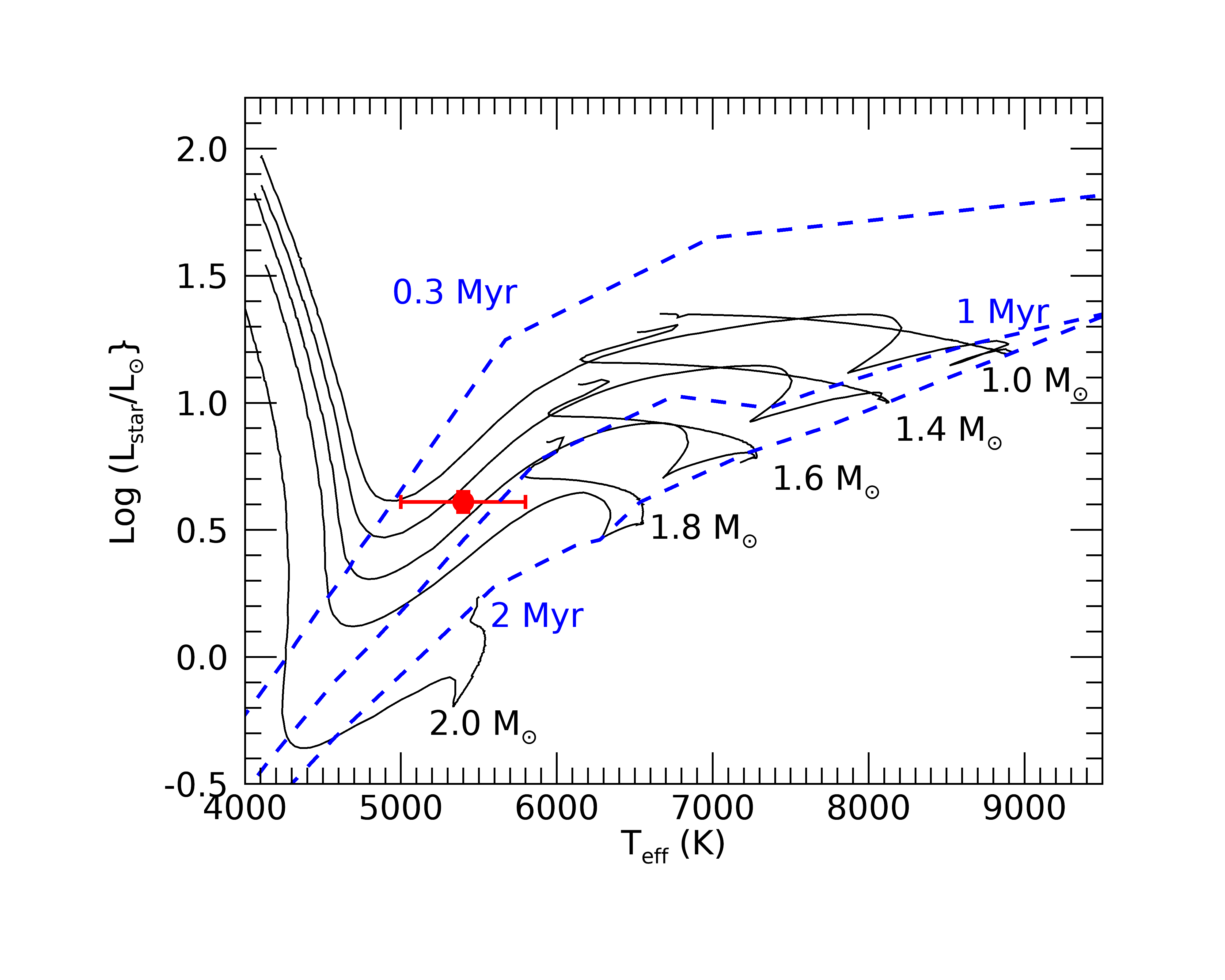}
\end{center}
\caption{Evolutionary tracks of Siess et al.\,(2000) in the range 1-2\,\msun\, (black) and for ages between 0.3 $-$ 2 Myr (blue). The red point represents Gaia23bab.}
\label{fig:fig6}
\end{figure}

The stellar properties of Gaia23bab were derived through
the near-infrared photometry and extinction in the quiescent
phase. For a distance of 900\,pc and \av\,=\,8.0\,mag, the 
absolute $J$, $H$, and $K_s$ magnitudes are 1.97\,mag, 1.67\,mag, and 
1.49\,mag, and the intrinsic [$J-H$] and [$H-K_s$] colors are 0.30\,mag  and 0.18\,mag, respectively. Considering the tables of Pecaut \& Mamajek (2013) for
sources with ages between 5 and 30 Myr, the
spectral type is G3$-$K0, and the effective temperature \teff\,=\,5400$\pm$400\,K. In Figure\,\ref{fig:fig5} we plot the black-body function at T\,=\,5400\,K, extincted for A$_V$\,=8.0\,mag (quiescence) and 4.8/6.2\,mag (burst, see Sect.\,\ref{sec:sec4.1}), to show the consistency with the observed photometric data.
The bolometric
magnitude, \Mbol\,, can be computed as  \Mbol\,=m($J$)+5-5\,log$_{10}$\,$d$(pc)+BC$_J$, m($J$) being the intrinsic $J$ 
magnitude and BC$_J$ the
bolometric correction. For a G3$-$K0 star BC$_J$ is 1.08$-$1.30. 
Therefore, we get  \Mbol\,
between 3.05 mag and 3.27 mag. Then, 
an estimate of the stellar 
luminosity, \lstar\,, can be 
obtained as 
log$_{10}$\,\lstar\,=\,0.4[$M_{\mathrm{bol},\odot}$-\Mbol\,], where $M_{\mathrm{bol},\odot}$ is the 
bolometric luminosity of the Sun, equal to
4.74 mag (Mamajek et al. 2015). This way, we get \lstar\,=\,4.0$\pm$0.5 \lsun\,. Therefore, having estimated as $\sim$ 5 \lsun\, the bolometric luminosity in quiescence  (Sect.\,\ref{sec:sec3.3}) we obtain \lacc$^{quiesc}$\,$\sim$\,1 \lsun\,. 
Assuming black-body emission, we determined the stellar radius, 
\rstar\, =\,1/2\teff$^2/\sqrt{L_{\mathrm{*}}/\pi\sigma}$, $\sigma$ being the Stefan-Boltzmann constant.
We find  \rstar = 2.2$\pm$0.5 \rsun.

An estimate of the evolutionary status of Gaia23bab is determined from the spectral index $\alpha = \frac{d\log (F_\lambda \lambda)}{d\log (\lambda)}$ between 2.16\,$\mu$m and 22\,$\mu$m (WISE) or 24\,$\mu$m (Spitzer-MIPS). In quiescence we get $\alpha$\,=$-$0.52, while in burst $\alpha$\,=$-$0.77. Both these determinations are in agreement with the spectral index  $\alpha$\,$<-$0.3 typical of Class II sources (Greene \& Lada 1996). Also the Spitzer colors, ([3.6]$-$[4.5])\,=\,0.5 and ([4.5]$-$[5.8])\,=\,0.77, are in the range predicted for Class II sources by the models of Allen et al. (2004) and Megeath et al. (2004), namely
0.0 $<$([3.6]\,$-$\,[4.5]) $<$ 0.7 (0.8) and 0.4 $<$ ([5.8]\,$-$\,[8.0]) $<$ 1.0 (1.1).  An age $\la$\, 1 Myr can be estimated by comparing \lstar\, and \teff\, with the evolutionary models of Siess et al. (2000, Figure\,\ref{fig:fig6}), where the location of Gaia23bab is consistent with \mstar\,=\, 1.6$\pm$0.1 \msun.

\section{Spectroscopic analysis}\label{sec:sec4}
\begin{figure*}
\begin{center}
\includegraphics[width=0.8\columnwidth]{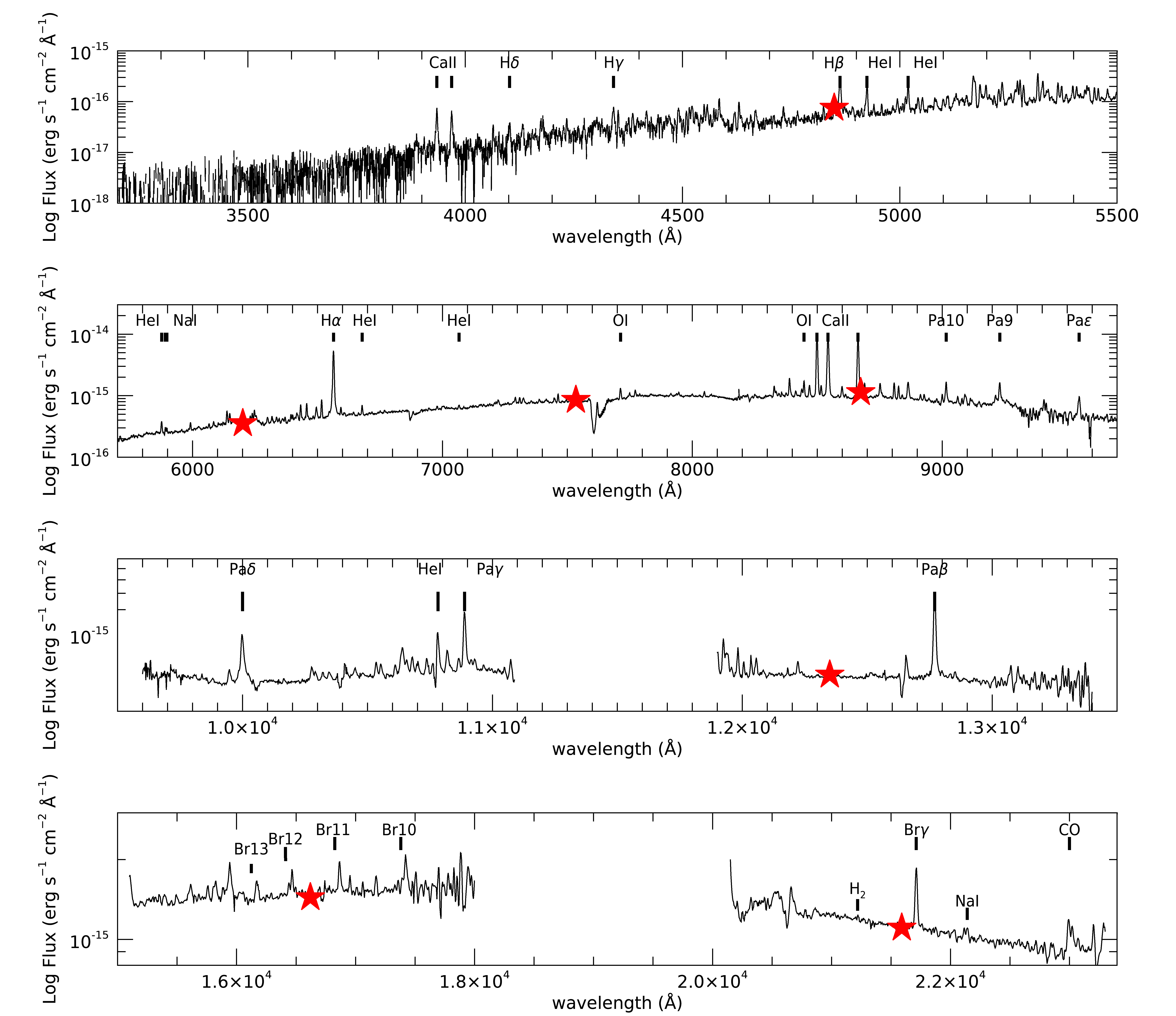}
\caption{
 MODS and LUCI spectra of Gaia23bab. The photometric points in $griz$ and $JHK_s$ bands are shown with asterisks. The main spectroscopic emission lines are labeled.
\label{fig:fig7}}
\end{center}
\end{figure*}

MODS and LUCI spectra of Gaia23bab are presented 
in Figure\,\ref{fig:fig7}. The spectrum steeply rises between 3500 \AA\, and 8000 \AA\, likely because of the relevant extinction that affects Gaia23bab (Sect.\,\ref{sec:sec3.2} and \ref{sec:sec4.1}). As a consequence, in this wavelength range the spectrum is also very noisy, therefore preventing the possible detection of the continuum excess emission (Balmer jump) between 3600 and 4000 \AA\,, which represents a direct signature of accretion (e.g. Alcal\'a et al. 2014 and references therein). At wavelengths between 0.9 $\mu$m and 1.6 $\mu$m the spectrum is almost flat and then decreases at wavelengths longer than 2 $\mu$m, in agreement with what expected for Class II sources. 
A number of emission lines are detected, and in particular those originating in the accretion columns,
such as \hi\, and \hei\, 
recombination lines,  bright \caii\, and \oi\, lines, and 
weaker metallic lines (Alcal{\'a} et al. 2014, and references therein). In addition, emission from the disk 
(CO 2-0 bandhead, and \nai\, doublet at 2.2 $\mu$m) along with 
weak emission from outflowing gas (H$_2$ 1-0\,S(1) 2.12\,$\mu$m and \hei\, 1.08\,$\mu$m, which presents also a blue-shifted 
absorption component), is detected. No forbidden atomic lines 
are present in the spectrum. Line fluxes and their 
1$\sigma$ errors were computed using the {\rm SPLOT} task in {\rm IRAF}, which takes into account both the effective readout noise per pixel and the photon noise in the spectral region containing the emission line\footnote{The effective readout noise per pixel was measured as the root mean square deviation ({\it rms}) of the continuum in either side of each line. The average {\it rms } was then set as  parameter '$\sigma_0$' in SPLOT. The photon noise was estimated as {\it invgain*I}, where {\it invgain} is the reciprocal of the MODS/LUCI gain (2.5/2.0 e$^{-}$/ADU expressed in physical units) and  {\it I} is the pixel value.  The error on the profile fit is then computed  by a Monte-Carlo simulation, whose iteration number {\it nerrsample} was set  to 100.}. 
Fluxes of the main emission lines are given in Table\,\ref{tab:tab2}.

\begin{table}
\center
\caption{\label{tab:tab2} Fluxes of the main observed lines}
\begin{tabular}{ccc}
\hline
\hline
Line & $\lambda$ & F$\pm\Delta$F \\
    & (\AA)& 10$^{-16}$ erg s$^{-1}$ cm$^{-2}$) \\
%\decimalcolnumbers
\cline{1-3}
CaII H       & 3934  & 2.6$\pm$0.4 \\
CaII K       & 3968  & 2.3$\pm$0.4\\
H$\delta$    & 4102  & 1.3$\pm$0.4 \\
H$\gamma$    & 4340  & 2.6$\pm$0.3\\
H$\beta$     & 4861  &10.5$\pm$0.6 \\
He I         & 4922  & 5.9$\pm$0.3 \\
He I         & 5015  & 5.6$\pm$0.3 \\
He I         & 5875  & 7.0$\pm$1.0 \\
Na I         & 5890  & 2.3$\pm$0.9 \\
Na I         & 5896  & 2.6$\pm$0.9 \\
H$\alpha$    & 6562  & 299.0$\pm$1.8 \\
He I         & 7065  & 6.1$\pm$2.2  \\
O I          & 7774  & 25.1$\pm$1.6 \\
O I          & 8446  & 37.6$\pm$1.7 \\
Ca II        & 8498  & 538.9$\pm$1.9\\
Ca II        & 8542  & 590.9$\pm$1.8\\
Ca II        & 8662  & 507.8$\pm$1.8\\
Pa10         & 9015  & 59.2$\pm$1.8 \\
Pa9          & 9229  & 79.1$\pm$2.3 \\
Pa8          & 9545  & 54.3$\pm$1.8 \\
Pa$\delta$   & 10052 & 230.0$\pm$5.2\\
He I$^a$     & 10833 & (-26.1)122.3$\pm$4.5 \\
Pa$\gamma$   & 10941 & 244.2$\pm$4.9  \\
Pa$\beta$    & 12821 & 380.4$\pm$4.5   \\
Br13         & 16114 & 70.3$\pm$5.2    \\
Br12         & 16412 & 100.2$\pm$7.4   \\
Br11         & 16811 & 130.1$\pm$4.1  \\
Br10         & 17367 & 177.2$\pm$6.9   \\
H$_2$ 1-0S(1)& 21218 & 9.0$\pm$4.0    \\
Br$\gamma$   & 21661 & 190.0$\pm$4.0   \\
Na I         & 22062 & 18.2$\pm$6.0    \\
Na I         & 22090 & 23.2$\pm$6.0    \\
CO 2-0       & 22992 & 184.0$\pm$8.8  \\ 
\hline
\end{tabular}
\tablecomments{The line wavelength is in air/vacuum for $\lambda$ $<$/$>$ 1\,$\mu$m, respectively.
$^a$In parenthesis we give the flux of the absorption component of the line.}
\end{table}

\subsection{Extinction during burst}\label{sec:sec4.1} 
\begin{figure}
\begin{center}
\includegraphics[width=0.8\columnwidth]{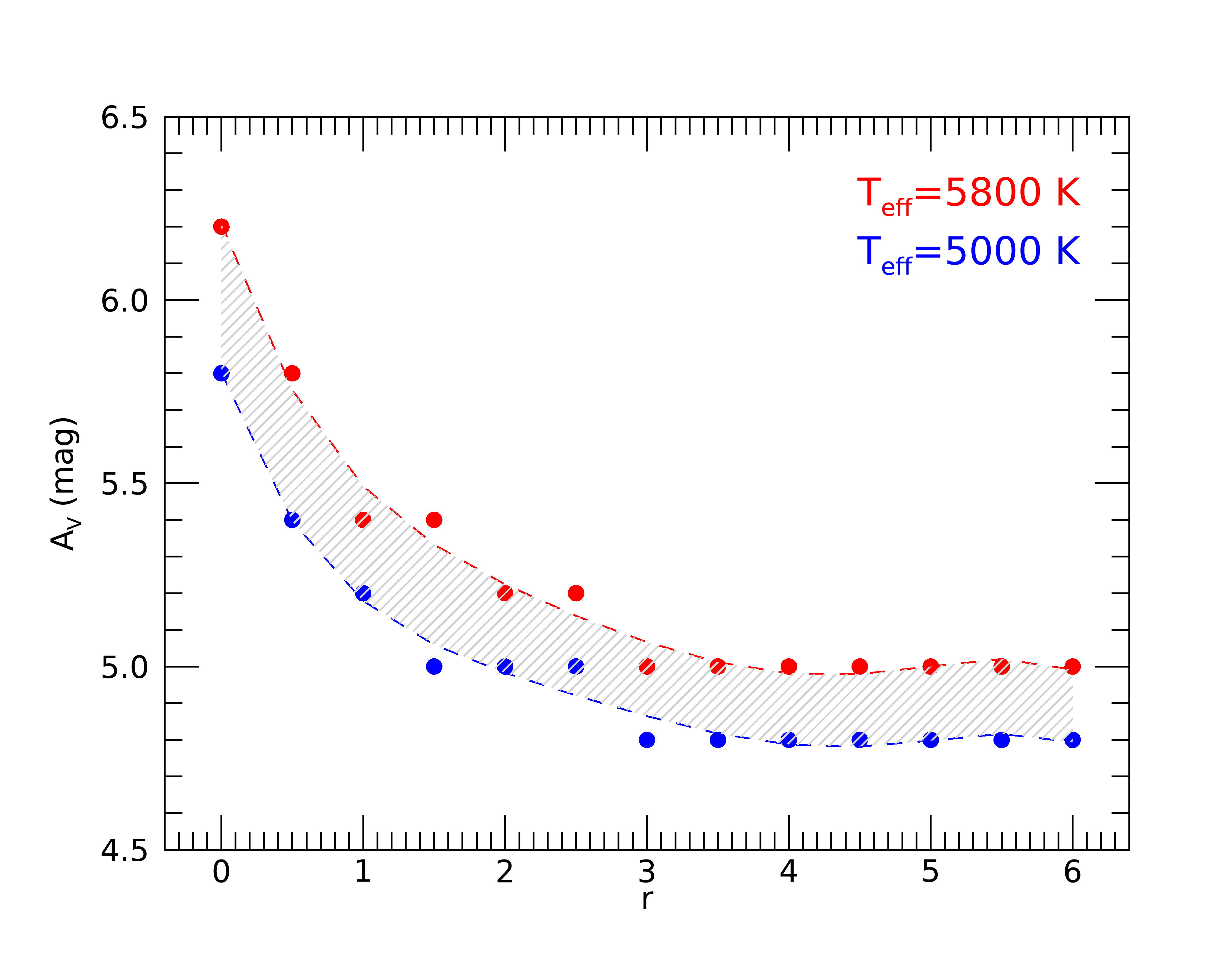}
\end{center}
\caption{\label{fig:fig8}\av\, vs. $r$ (assumed constant)
for a stellar effective temperature \teff\, between 5000 K and 5800 K (see Sect.\ref{sec:sec4.1}).} 
\end{figure}

A first \av\, estimate of 5$-$6 mag for the 2023 burst has been derived from the optical/near-infrared color-color diagrams (Sect.\ref{sec:sec3.2}).  
An independent measure is based on the ratio of the  observed continuum with that of a stellar template of the same spectral type as the target, artificially reddened by varying the value of \av\,(Alcal{\'a} et al. 2021). The best estimate of \av\ is obtained when this ratio has a flat slope.
Although this method is commonly used in classical T Tauri stars, in strong accretors the observed continuum is not only affected by extinction, but can be also significantly enhanced by the excess continuum coming from hot spots in the accretion shock and from the disk. Such excess is described in terms of veiling, $r$\,=\,Flux(excess)/Flux(star), usually estimated by comparing the equivalent width of the photospheric lines in the template spectrum with those in the target spectrum. In Gaia23bab, however, no photospheric lines are detected, either because deep bands/lines are not expected in spectra of G$-$K type stars (Herczeg \& Hillenbrand 2014), or because of a high veiling.  We can consider a reasonable range of $r$ basing from literature estimates, however. In classical T Tauri stars, between 6000$-$8000\,\AA\,  $r$ is typically between 0 and 2 (Fischer et al. 2011, Alcal{\'a} et al. 2021),  but it can be $\ga$ 3 in strong accretors (e.g. Giannini et al. 2022). In this wavelength range $r$ is  roughly constant or decreases as a black-body at $T$\,$\approx$\,8\,000\,K (considered as the temperature of the hot spot giving rise to the optical veiling, Fischer et al. 2011). To apply the procedure described above to compute \av\,, we adopted as a first approximation a constant $r$ between 0 and 6  and added the relative excess continuum  to the stellar template prior than applying the variable reddening.
We used the optical  templates for stars with \teff\,=\,5000$-$5800\,K  of Gonneau et al. (2020), and considered  a grid of \av\, between 0 and 20\,mag, adopting the extinction law of Cardelli et al. (1989) and a total-to-selective extinction ratio R$_V$\,=\,3.1. This way, we estimated \av\, as a function of $r$, as shown in Figure\,\ref{fig:fig8}.
We note that for small values of
$r$ ($r$\,$\la$ 3), \av\, decreases with $r$\, as expected if we consider that extinction effects tend to diminish the continuum, while veiling acts in the opposite direction. As $r$\, increases, and in particular when the intensity of the excess continuum prevails over the stellar one, the spectral shape 
no longer changes, and \av\, tends to an asymptotic minimum. 
From this plot we derive \av\,=\,5.5$\pm$0.7 mag. Finally, we have also checked how the derived \av\, values change assuming that the veiling follows a black-body law at T\,=\,8000 K. The results do not change significantly with respect to the case of a constant veiling.

\subsection{Accretion properties}\label{sec:sec4.2}
A rough measure of the accretion luminosity due to the 2023 burst event can be derived as \lbol$^{burst}$\,-\,\lstar\,=\,6.3$\pm$3.9\,\lsun\,.
A more accurate method relies on the empirical relationships found by Alcal{\'a} et al. (2014, 2017), between the accretion luminosity, \lacc, and the luminosities \lumi\, of 
selected emission lines of gas in the accretion columns.
In the optical range these relationships exist for more than 20 lines, namely the \hi\, recombination lines of the Balmer and 
Paschen series from H$\alpha$ to H15, and from Pa8 to Pa10, along with 
\hei\,, \oi\, and \caii\, lines. In the near-infrared, there are relationships for Pa$\delta$, Pa$\gamma$, Pa$\beta$, and Br$\gamma$.
In the Gaia23bab spectrum there are 24 lines useful for determining \lacc\,. For a fixed \av\,, the best estimate of \lacc\, is that for which the dispersion among the individual \lacci\, is minimized. The associated error is the combination of the uncertainties on the line fluxes and those on the relationships between \lacc\ and \lumi\, (Alcal{\'a} et al. 2017). Typically, this error does not exceed a few tenths of solar luminosity, negligible with respect to that induced by the uncertainty on \av\,. For  \av\,=\,5.5$\pm$0.7 mag we obtain \lacc$^{burst}$\,=\,3.7$\pm$1.8 \lsun\,.

From \lacc$^{burst}$\,, \mstar\,, and \rstar\,, we derived the mass accretion rate during the burst as 
\macc$^{burst}$ =  (1 - \rstar/R$_{\rm in}$)$^{-1}$ \lacc$^{burst}$ \rstar/G \mstar\, (Gullbring et al. 1998),
where R$_{\rm in}$ is the inner-disk radius, assumed  $\sim$ 5\rstar\, (Hartmann et al. 1998), and G is the gravitational constant.
We obtain \macc$^{burst}$\,=\,(2.0$\pm$1.0)\,10$^{-7}\,$\msunyr\,, in line with the mass accretion rate values found in classical EXor events (e.g. Audard et al. 2014). From the same relation, and adopting the value of \lacc$^{quiesc}$\, derived in Sect.\,\ref{sec:sec3.4}, we obtain \macc$^{quiesc}$\,$\sim$ 6\,10$^{-8}\,$\msunyr\,. All the stellar and accretion parameters are summarized in Table\,\ref{tab:tab3}.

\begin{table*}
\center
\caption{\label{tab:tab3} Stellar and accretion properties}
\begin{tabular}{cc|ccc}
\hline
\hline
\multicolumn{2}{c}{Stellar properties} & & \multicolumn{2}{c}{Accretion properties}\\
\hline
\lstar (\lsun) &  4.0$\pm$0.5   &                 & Quiescence   &   Outburst    \\
\cline{4-5}
\mstar (\msun) &  1.6$\pm$0.1  & \lbol (\lsun)   & $\sim$ 5      &   10.3$\pm$3.9       \\
\rstar (\rsun) & 2.2$\pm$0.5   & \av (mag)       & $\sim$ 8      &   5.5$\pm$0.7        \\
\teff (K)      &  5400$\pm$400 & \lacc (\lsun)   & $\sim$ 1      &   3.7$\pm$1.8 \\
SpT            &  G3$-$K0         & \macc (\msunyr) & $\sim$ 6 10$^{-8}$ &(2.0$\pm$1.0) 10$^{-7}$\\
%$\alpha$      & -0.52$-$-0.77 
\hline\end{tabular}	
\end{table*}

\section{Discussion}\label{sec:sec5}
\subsection{The role of bursts in assembling mass}\label{sec:sec5.1}
One of the fundamental questions about EYSs is the role of bursts for the stellar mass assembly (Fischer et al. 2023). We have estimated that during the 2023 burst, the luminosity due to accretion was comparable to the stellar luminosity (Table\,\ref{tab:tab3}).  
In the last 10 years Gaia23bab has undergone three bursts, lasting approximately one year each  (Table\,\ref{tab:tab1}). 
Furthermore, they have similar amplitude and all present a 'triangular' shape in the light curve. This allows us to assume as the average value of the mass accretion rate throughout each event, half the value of \macc$^{burst}$ measured close to the peak of the 2023 burst.
We obtain that over the last ten years, three of which were spent in burst, the mass accumulated on Gaia23bab was $\sim$ 1\,10$^{-6}$\,\msun\,, namely roughly twice the mass that the star would have assembled if it had remained quiescent for the same period of time. Notably, this estimate is quite similar to that evaluated for the 2022 burst of EX Lupi (Cruz-S{\'a}enz de Miera et al. 2023). 
Considering that the burst amplitude is thought to decrease with time (e.g. Fischer et al. 2023), we speculate that the contribution of burst episodes to the final stellar mass could be even higher than our estimate.

\subsection{Gaia23bab in the context of EXors}\label{sec:sec5.2}

\begin{figure}[ht!]
\begin{center}
\includegraphics[width=1\columnwidth]{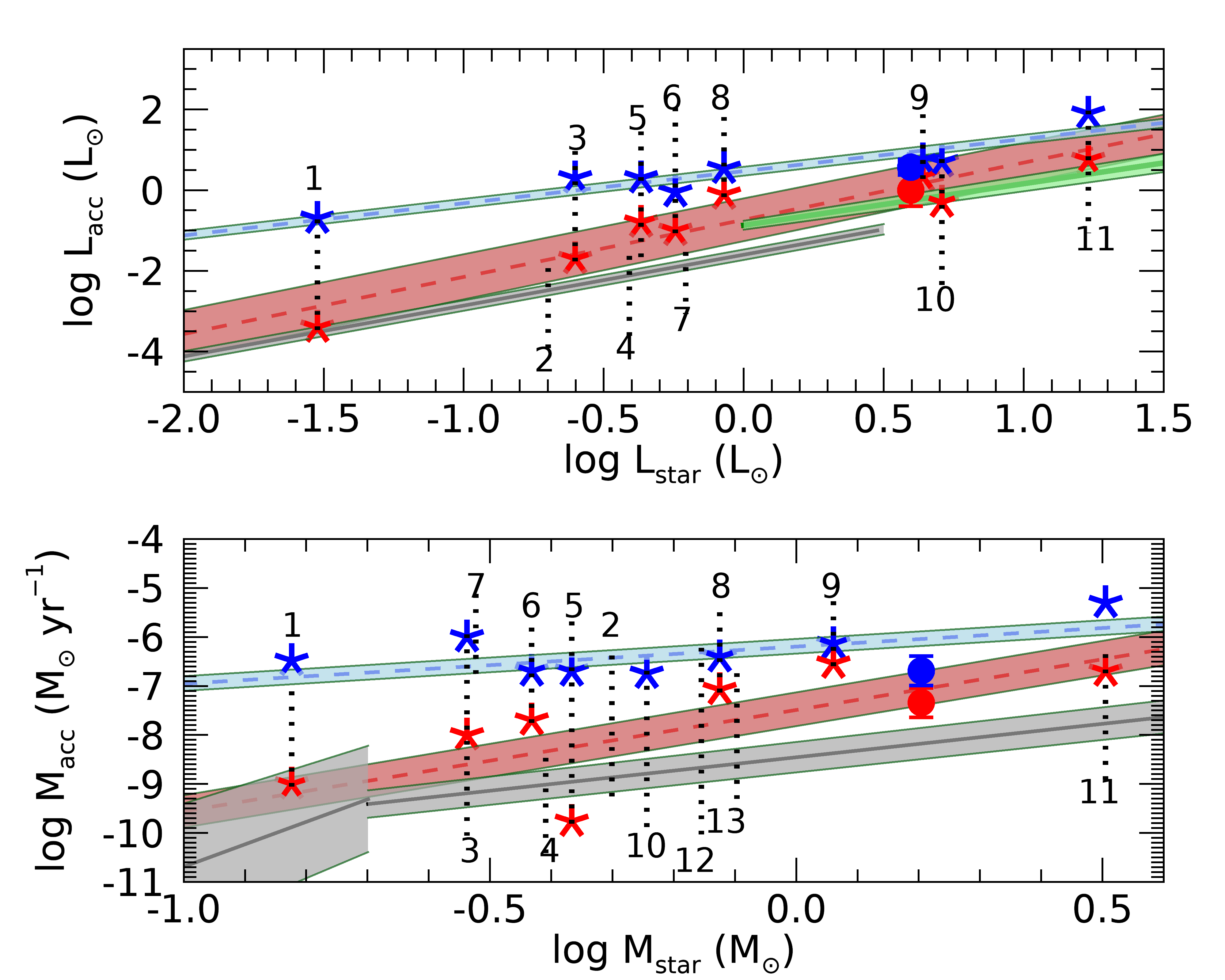}
\end{center}
\caption{
\label{fig:fig9}  \lacc\, vs. \lstar\, (top panel) and \macc\, vs. \mstar\, (bottom panel) in a sample of known EXors. Individual sources are identified as following : \#1:\,ASASSN-13db ([1,2]); \#2:\,V1143 Ori ([3,4]); \#3:\,V1118 Ori ([5,6]); \#4:\,VY Tau ([3,4]);  \#5:\,EX Lup ([7,8], note that for this source the determinations of the accretion parameters for two bursts are available); \#6:\,XZ Tau ([3,9,10,11]);  \#7:\,UZ Tau E ([3,12]);  \#8:\,DR Tau ([11,13]);  \#9:\,Gaia20eae ([14]);  \#10:\,Gaia19fct ([3,15]); \#11:\,PV Cep ([3 and references therein]); \#12:\,V2492 Cyg ([16]);  \#13:\,V1180 Cas ([17]).  Red and blue asterisks are \lacc\, (\macc)\, in quiescence and in burst, respectively. The filled circles represent Gaia23bab. The red and blue dotted lines are the best linear fit through the data in quiescence and in burst and the shaded areas the correspondent 1\,$\sigma$\, standard deviation. The grey/green solid lines are the relationships valid for classical T Tauri/low-mass HAeBe stars, and the shaded areas are the associated uncertainty (Alcal\'a et al. 2017, Wichittanakom et al. 2020). References to the figure:[1]\,-\,Holoien et 
al.\,2014; [2]\,-\,Sicilia-Aguilar et al.\,2017;
[3]\,-\,Giannini et al.\,2022; [4]\,-\,Sipos \& 
K{\'o}sp{\'a}l\,2014; [5]\,-\,Giannini et 
al.\,2017; [6]\,-\,Giannini et al.\,2020; [7]\,-\,Aspin et al.\,2010;
[8]\,-\,Cruz-S{\'a}enz de Miera et al.\,2023; [9]\,-\,Osorio et al.\,2016; [10]\,-\,Hartigan \& 
Kenyon\,2003;  [11]\,-\,Antoniucci et al.\,2017; [12]\,-\,Yang et al.\,2012; [13]\,-\,Banzatti et 
al.\,2014; [14]\,-\,Cruz-S{\'a}enz de Miera et 
al.\,2022; [15]\,-\,Park et al.\,2022; [16]\,-\,Giannini et al.\,2018; [17]\,-\,Kun et al.\,2011.
}
\end{figure}

In the previous sections, we have analysed the photometric features of Gaia23bab to show their similarities with EXor sources. Here, we focus on the parameters derived from the emission lines, namely the accretion luminosity and the mass accretion rate. In Figure\,\ref{fig:fig9} we plot \lacc\, vs. \lstar\, and \macc\, vs. \mstar\, for a sample of 13 known EXors, both in quiescence and in burst. For comparison, the {\it loci} of accreting T Tauri stars (Alcal\'a et al. 2017) and of low-mass HAeBe (Wichittanakom et al. 2020) stars, are shown with a grey and green shaded areas, respectively. 
For the large majority of the sources, \lacc\, was measured in the same way, namely applying the Alcal\'a et al. (2017) relations between  optical/NIR emission lines and accretion luminosity. Both in quiescence and in burst we find a tight relation between \lacc\, and \lstar\,, that spans more than two orders of magnitude in \lstar\,. A good correlation is also found between \macc\, and \mstar\, but with a larger spread that is likely due to the uncertainty in the stellar mass and radius determinations. The linear fits through the data points are:
\begin{equation}
  {\rm log}\, {L}^{quiesc}_{\mathrm{acc}}\,=\,(1.41 \pm 0.21)\, {\rm log}\, {L}_{\mathrm{*}}\, - (0.73 \pm 0.17) 
\end{equation}
\begin{equation}
  {\rm log}\, \dot{M}^{quiesc}_{\mathrm{acc}}\,=\,(2.07 \pm 0.77)\, {\rm log}\, {M}_{\mathrm{*}} - (7.49 \pm 0.33) 
 \end{equation}
in quiescence, and
\begin{equation}
  {\rm log}\, {L}^{burst}_{\mathrm{acc}}\,=\,(0.79 \pm 0.14)\,{\rm log}\, {L}_{\mathrm{*}}\, + (0.14 \pm 0.11)  
  \end{equation}
 \begin{equation}
 {\rm log}\, \dot{M}^{burst}_{\mathrm{acc}}\,=\,(0.75 \pm 0.37)\,  {\rm log}\, {M}_{\mathrm{*}} - (6.19 \pm 0.15) 
\end{equation}
in burst.\\
The values measured in Gaia23bab during the burst are in agreement with both fits, if we take into account that our spectra have been taken at the beginning of the declining phase, when the $r$ magnitude was already increased by about 0.5 mag with respect to the peak traced by the ZTF data in the same band (Sect.\ref{sec:sec3.1}). Instead, the accretion properties estimated in quiescence are consistent with the expected values. We recall that they have been indirectly derived from the SED, and thus a quiescent spectrum would be necessary to derive accurate determinations of \lacc$^{quiesc}$\, and \macc$^{quiesc}$\,. Also, this will allow us to investigate whether a variable extinction has a role in the different brightness phases.\\
More in general, we note that: 1) the angular coefficients of the relations derived in quiescence are  consistent within the uncertainties with those of T Tauri stars. Instead, the intercepts differ by about an order of magnitude. If confirmed on the base of a larger sample, this result would imply that EXors, even in quiescence, are more efficient than T Tauri stars in accreting mass; 2) when in burst, the relations between accretion and stellar parameters become shallower, as if there were a limit to the amount of material that can be transferred from the disk to the star through the accretion columns. Given the low number of sources for which the accretion vs. stellar properties have been derived, and the fact that the high-mass regime is represented only by PV Cep data, these results need to be confirmed on a larger statistical base, however.

\section{Summary}\label{sec:sec6}
In this paper we have presented LBT observations of the $\sim$ 2 mag burst of Gaia23bab, a YSO alerted by Gaia on March 2023. Our results, derived from the analysis of photometric and spectroscopic data, can be summarized as follows:
\begin{itemize}
    \item[1.] The multi-wavelength light curve shows that Gaia23bab had three bursts in the last ten years. These bursts are quite similar to each other in amplitude, duration, and rising/declining speed, in line with those observed in EXors. 
    \item[2.] We determined the stellar properties of Gaia23bab. It is a 1.6 \msun\,, Class II source with age $\la$\, 1 Myr, spectral type G3$-$K0 and stellar luminosity 4.0 \lsun\,. The accretion luminosity and the mass accretion rate in quiescence are \lacc$^{quiesc}$\,$\sim$\,1\,\lsun, and \macc$^{quiesc}$\,$\sim$\,6 10$^{-8}$ \msunyr. 
    \item[3.] The optical/NIR spectrum is rich in emission lines from which we have measured the accretion luminosity and the mass accretion rate during the burst. We get \lacc$^{burst}$\,$\sim$\,3.7\, \lsun, comparable with the stellar luminosity.
    The mass accretion rate close to the burst peak is \macc$^{burst}$\,$\sim$\,2.0 10$^{-7}$ \msunyr. 
    \item[4.] The mass accumulated on Gaia23bab in the last ten years  was roughly twice the mass that it would have assembled if it had remained quiescent for the same period of time.
    \item[5.] Both accretion luminosity and mass accretion rate of Gaia23bab are consistent with those of confirmed EXors. 
\end{itemize}
As a more general result, we have quantified the correlations, both in quiescence and in burst, between accretion and stellar parameters in a sample of 13 EXors. On average, EXors have \lacc\, and \macc\, larger than T Tauri stars in the same range of mass, even in quiescence. If confirmed on the base of a larger sample, this result would imply that EXors are more efficient than T Tauri stars in accreting mass. Also, when in burst, accretion luminosity and mass accretion rate become poorly dependent on the central star properties.

\begin{acknowledgments}
\section{Acknowledgements}
This work is based on observations made with the Large Binocular Telescope (LBT programme IT2023B-005). The LBT is an international collaboration among institutions in the United States, Italy and Germany. LBT Corporation partners are: The University of Arizona on behalf of the Arizona university system; Istituto Nazionale di Astrofisica, Italy; LBT Beteiligungsgesellschaft, Germany, representing the Max-Planck Society, the Astrophysical Institute Potsdam, and Heidelberg University; The Ohio State University, and The Research Corporation, on behalf of The University of Notre Dame, University of Minnesota and University of Virginia. This work has been supported by the Large Grant INAF 2022 YODA (YSOs Outflows, Disks and Accretion: towards a global framework for the evolution of planet forming systems). We acknowledge support from the ESA PRODEX contract nr. 4000132054. We acknowledge the Hungarian National Research, Development and Innovation Office grant OTKA FK 146023.
Z.N. was supported by the J\'anos Bolyai Research Scholarship of the Hungarian Academy of Sciences. T.G.  acknowledges the INAF grant 'EXORCISM : photometric and spectroscopic characterization of young eruptive variables'. FCSM received financial support from the European Research Council (ERC)
under the European Union’s Horizon 2020 research and innovation
programme (ERC Starting Grant “Chemtrip”, grant agreement No 949278). This research has made use of the Spanish Virtual Observatory (https://svo.cab.inta-csic.es) project funded by MCIN/AEI/10.13039/501100011033/ through grant PID2020-112949GB-I00..   
\end{acknowledgments}

\vspace{5mm}
\facilities{LBT(MODS,LUCI), Gaia}

\end{document}